\documentclass[twocolumn,prb]{revtex4}
\usepackage{amsfonts}
\usepackage[T1]{fontenc}
\usepackage{amsmath,amsbsy,amssymb,graphicx}
\usepackage{times}

\begin{document}

\title{Topological parafermion corner states \\
in clock-symmetric non-Hermitian second-order topological insulator}
\author{Motohiko Ezawa}
\affiliation{Department of Applied Physics, University of Tokyo, Hongo 7-3-1, 113-8656,
Japan}

\begin{abstract}
Parafermions are a natural generalization of Majorana fermions. We consider
a breathing Kagome lattice with complex hoppings by imposing $\mathbb{Z}_{3}$
clock symmetry in the complex energy plane. It is a non-Hermitian
generalization of the second-order topological insulator characterized by
the emergence of topological corner states. We demonstrate that the
topological corner states are parafermions in the present $\mathbb{Z}_{3}$
clock-symmetric model. It is also shown that the model is realized in
electric circuits properly designed, where the parafermion corner states are
observed by impedance resonance. We also construct $\mathbb{Z}_{4}$ and $%
\mathbb{Z}_{6}$ parafermions on breathing square and honeycomb lattices,
respectively.
\end{abstract}

\maketitle

\section{Introduction}

Topological quantum computation is a fault-tolerant quantum computation\cite%
{Moore,Das,Kitaev,TQC,SternA,Stern,NPJ}. Majorana fermions provide us with a
most studied platform of topological quantum computation\cite%
{Been,Stan,Elli,AA,Ivanov,Halperin}. The Majorana operator $\gamma $
satisfies $\gamma ^{2}=1$. They are realized as topological boundary states
of topological superconductors\cite{AliceaBraid,Qi,Lei,Alicea,Sato,Tanaka}
and Kitaev spin liquids\cite{Kitaev,Matsuda}. However, it is impossible to
perform universal quantum computation only by the braiding of Majorana
fermions since they can reproduce only a part of Clifford gates\cite{Ahl}.

Parafermions are straightforward generalization of Majorana fermions\cite%
{FendJSM,Fend,AliceaPara,Jerm,Ebisu}, where the parafermion operator $\gamma 
$ satisfies $\gamma ^{d}=1$ for $d\geq 3$. Braiding of parafermions with $%
d=3 $ are known to reproduce all the Clifford gates\cite{Hutter} although
universal quantum computation is not yet possible. In this sense,
parafermions are more powerful than Majorana fermions in the context of
quantum computation. Parafermions are realized in clock-spin models\cite%
{Baxter,Fend}, fractional quantum Hall effects\cite{ReadR,Clarke},
fractional topological superconductors\cite{Laub1} and twisted bilayer
graphene\cite{Laub2}. Among them, the $\mathbb{Z}_{d}$ clock-spin model\cite%
{Baxter} is non-Hermitian and its energy spectrum is $\mathbb{Z}_{d}$
symmetric in the complex plane. It is an interesting problem if they also
emerge as topological boundary states in certain lattice structures just as
Majorana fermions do.

Higher-order topological insulators and superconductors are generalization
of topological insulators and superconductors\cite%
{Fan,Science,APS,Peng,Lang,Song,Bena,Schin,FuRot,Kagome,EzawaPhos,Gei,Kha,EzawaMajo}%
. They are prominent by the emergence of zero-energy corner states instead
of gapless edge states. These zero-energy corner states are topologically
protected. A typical example is given by the breathing Kagome lattice\cite%
{Kagome}, where three topological corner states emerge. There are some
generalization to non-Hermitian higher-order topological insulators\cite%
{LiuSOTI,EzawaLCR,EzawaSkin,Berg}.

In this paper, generalizing the breathing Kagome second-order topological
insulator model by imposing $\mathbb{Z}_{3}$ clock symmetry, we propose a
new type of non-Hermitian higher-order topological insulator, where the
topological corner states are parafermions. This model is non-Hermitian,
where the energy spectrum is $\mathbb{Z}_{3}$ symmetric in the complex
energy plane as in the case of the $\mathbb{Z}_{3}$ clock-spin model. We
demonstrate how to implement the present model of parafermions in an
electric circuit. We also construct $\mathbb{Z}_{4}$ and $\mathbb{Z}_{6}$
parafermions as topological corner states on breathing square and honeycomb
lattices.

\section{Majorana fermion and parafermion}

Majorana fermion operators $\gamma _{i}$ satisfy the relations%
\begin{equation}
\left( \gamma _{j}\right) ^{2}=1,\qquad \gamma _{j}\gamma _{k}=-\gamma
_{k}\gamma _{j}.
\end{equation}%
Majorana fermions are realized as zero-energy states of a topological
superconductor, where particle-hole symmetry (PHS) preserves. It is
understood as follows. The PHS operator $\Xi $ acts as $\Xi ^{-1}H\Xi =-H$
with the eigen equation $H\left\vert \psi \right\rangle =E\left\vert \psi
\right\rangle $. If a particle has an energy $E$, its antiparticle has the
energy $-E$ in the presence of PHS. Namely, the wave functions always appear
in a particle-hole pair with a pair of energies $\left( E,-E\right) $. If
the states satisfy the relation $E=-E$, the particle is identical to its
antiparticle, and a pair of Majorana fermions emerge. Hence, the zero-energy
($E=0$) states respecting PHS are Majorana states.

Parafermions are natural generalization of Majorana fermions. $\mathbb{Z}%
_{d} $ parafermions are defined through the relations%
\begin{equation}
\left( \gamma _{j}\right) ^{d}=1,\qquad \gamma _{j}\gamma _{k}=\omega \gamma
_{k}\gamma _{j},  \label{Para}
\end{equation}%
where $\omega =e^{2\pi i/d}$. The minimal model consists of two elements $%
\gamma _{1}$ and $\gamma _{2}$.

\section{$\mathbb{Z}_{3}$ Parafermion}

We start with a minimal model by setting $d=3$ in Eq.(\ref{Para}). $\mathbb{Z%
}_{3}$ parafermions are represented by the shift operator\cite%
{Zohar,Fend,AliceaPara}%
\begin{equation}
\gamma _{1}\equiv \tau =\left( 
\begin{array}{ccc}
0 & 0 & 1 \\ 
1 & 0 & 0 \\ 
0 & 1 & 0%
\end{array}%
\right) ,  \label{Tau}
\end{equation}%
and the clock operator\cite{Zohar,Fend,AliceaPara}%
\begin{equation}
\gamma _{2}\equiv \sigma =\text{diag.}\left( 1,\omega ,\omega ^{2}\right) ,
\label{DigW}
\end{equation}%
where $\omega =e^{2\pi i/3}$. Here, $\tau $ and $\sigma $ satisfy the $%
\mathbb{Z}_{3}$ parafermion relations,%
\begin{equation}
\tau ^{3}=\sigma ^{3}=1,\qquad \sigma \tau =\omega \tau \sigma .
\label{EqPara}
\end{equation}%
In the $\mathbb{Z}_{3}$ clock-symmetric model, the energy spectrum is
composed of triplets\cite{Baxter} $E_{n}^{\left( 0,1,2\right) }$, $%
n=0,1,2,\cdots $,

\begin{equation}
E_{n}^{\left( 0,1,2\right) }=\varepsilon _{n},\quad \omega \varepsilon
_{n},\quad \omega ^{2}\varepsilon _{n},  \label{En}
\end{equation}%
satisfying $\varepsilon _{n}+\omega \varepsilon _{n}+\omega ^{2}\varepsilon
_{n}=0$. The system is necessarily non-Hermitian because the eigen energies
are complex except for zero-energy states.

\begin{figure}[t]
\centerline{\includegraphics[width=0.49\textwidth]{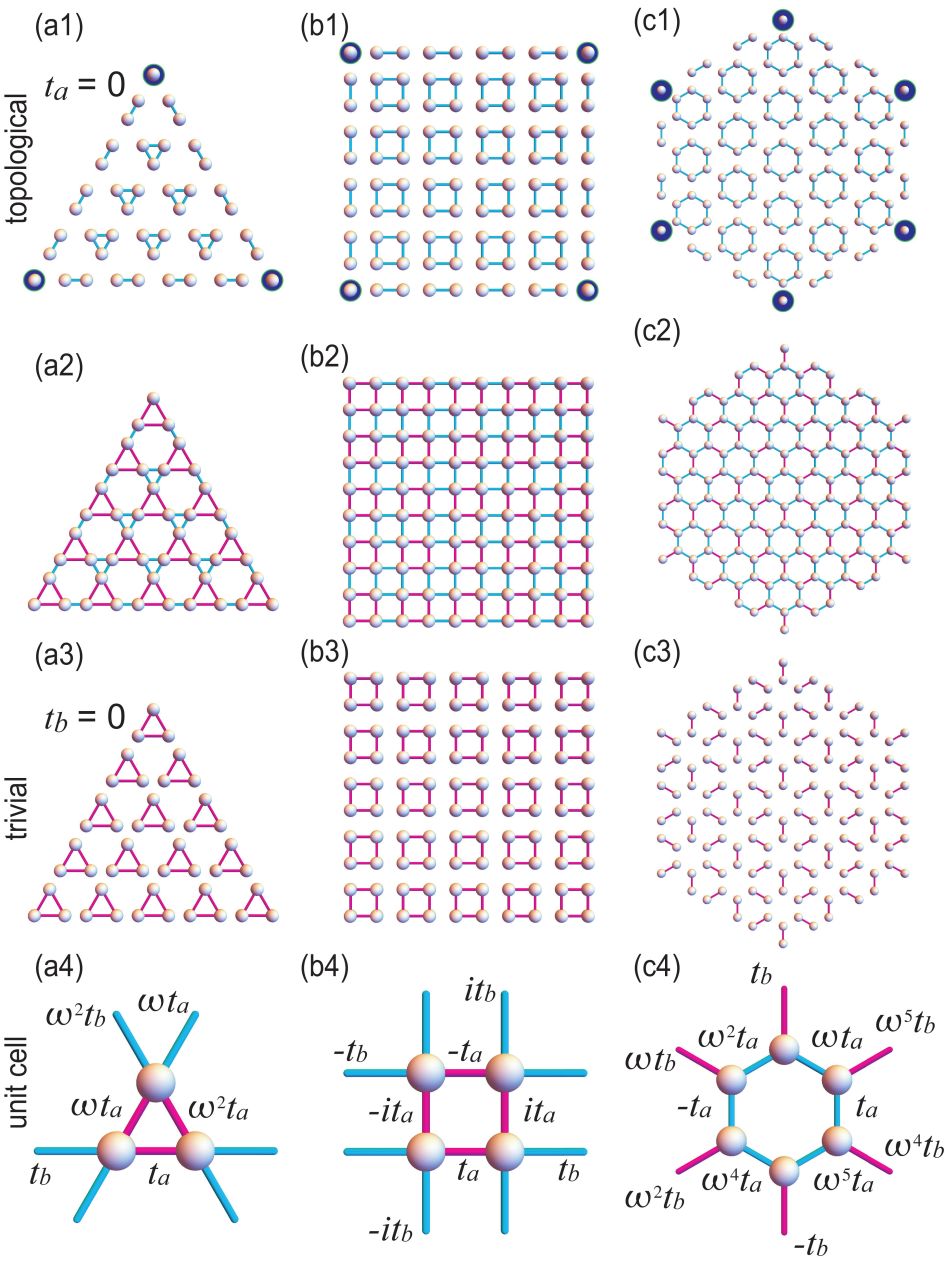}}
\caption{(a) Breathing Kagome lattice, (b) breathing square lattice and (c)
breathing honeycomb lattice. The hopping parameters are $t_{a}=0$ for (a1)$%
\sim$(c1), $t_{a}t_b\neq0$ for (a2)$\sim$(c2), and $t_{b}=0$ for (a3)$\sim$%
(c3). There emerge three, four and six parafermion corner states in the
topological phase. When $t_{a}=0$, they are isolated as marked by blue disks
in (a1), (b1) and (c1), respectively. The unit cells are given in (a4), (b4)
and (c4) with the hopping parameters indicated. }
\label{FigLattice}
\end{figure}

\subsection{Zero-energy parafermion states}

It follows from Eq.(\ref{En}) that only the zero-energy states form a set of
degenerate states respecting $\mathbb{Z}_{3}$ clock symmetry. They are $%
\mathbb{Z}_{3}$ parafermion states. We denote them as $\left\vert \psi
_{0}\right\rangle $, $\left\vert \psi _{1}\right\rangle $ and $\left\vert
\psi _{2}\right\rangle $. They are characterized by the properties 
\begin{align}
\tau \left\vert \psi _{0}\right\rangle & =\left\vert \psi _{1}\right\rangle
,\quad \tau \left\vert \psi _{1}\right\rangle =\left\vert \psi
_{2}\right\rangle ,\quad \tau \left\vert \psi _{2}\right\rangle =\left\vert
\psi _{0}\right\rangle ,  \label{TauP} \\
\sigma \left\vert \psi _{0}\right\rangle & =\left\vert \psi
_{0}\right\rangle ,\quad \sigma \left\vert \psi _{1}\right\rangle =\omega
\left\vert \psi _{1}\right\rangle ,\quad \sigma \left\vert \psi
_{2}\right\rangle =\omega ^{2}\left\vert \psi _{2}\right\rangle ,
\label{SigmaP}
\end{align}%
from which the matrix representations (\ref{Tau}) and (\ref{DigW}) follow.
Then, the $\mathbb{Z}_{3}$ parafermion relations (\ref{EqPara}) are verifed.
Namely, it is necessary and sufficient to examine Eqs.(\ref{TauP}) and (\ref%
{SigmaP}) for a triplet set of zero-energy states in order to show that they
are $\mathbb{Z}_{3}$ parafermions.

\subsection{Breathing Kagome lattice}

We propose a model possessing parafermions on the breathing Kagome lattice.
The bulk Hamiltonian is given by 
\begin{equation}
H=\left( 
\begin{array}{ccc}
0 & h_{12} & \omega h_{13} \\ 
h_{12}^{\ast } & 0 & \omega ^{2}h_{23} \\ 
\omega h_{13}^{\ast } & \omega ^{2}h_{23}^{\ast } & 0%
\end{array}%
\right) ,  \label{H3}
\end{equation}%
with%
\begin{align}
h_{12}& =t_{a}+t_{b}e^{ik_{x}}, \\
h_{23}& =t_{a}+t_{b}e^{-i\left( k_{x}/2+\sqrt{3}k_{y}/2\right) }, \\
h_{13}& =t_{a}+t_{b}e^{-i\left( k_{x}/2-\sqrt{3}k_{y}/2\right) },
\end{align}%
where we have introduced two hopping parameters $t_{a}$ and $t_{b}$,
corresponding to the magenta link and the cyan link along the horizontal
axis in Fig.\ref{FigLattice}(a4). The hopping parameters along the other two
triangle sides are given by $\omega t_{a}$ and $\omega ^{2}t_{a}$ for a
magenta triangle, and $\omega t_{b}$ and $\omega ^{2}t_{b}$ for a cyan
triangle. This model is non-Hermitian due to the presence of $\omega $.

A comment is in order with respect to the breathing Kagome model. It is a
typical model for the conventional second-order topological insulator\cite%
{Kagome}, where the factor $\omega $\ is absent and it is Hermitian. The
present generalization of the breathing Kagome lattice model provides us
with a new type of non-Hermitian second-order topological insulators.

\begin{figure}[t]
\centerline{\includegraphics[width=0.48\textwidth]{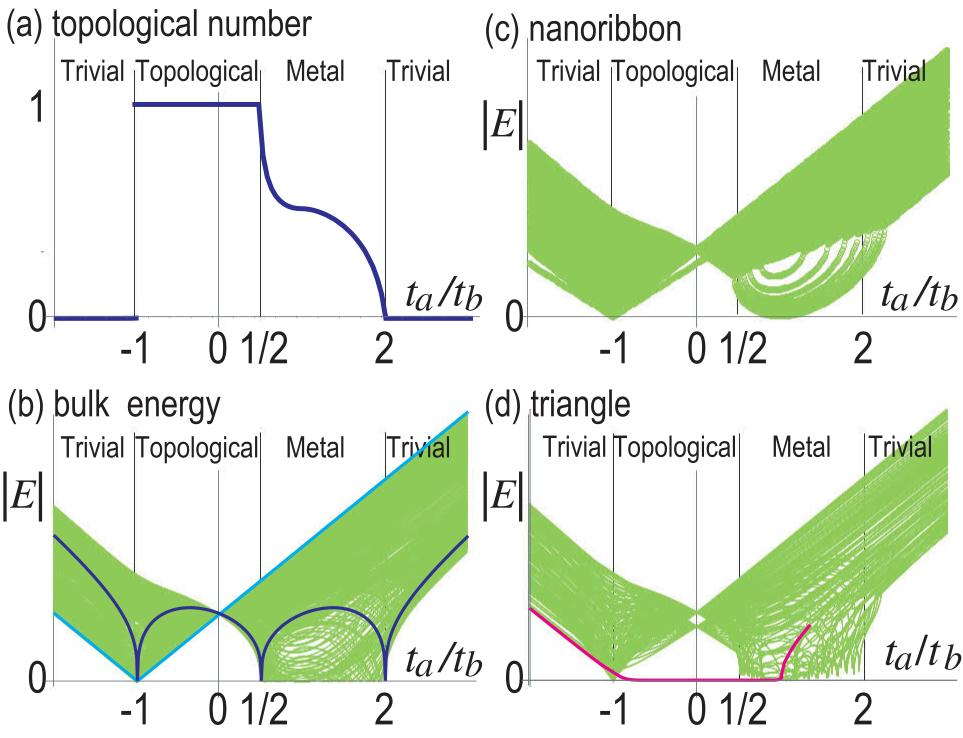}}
\caption{ (a) Topological number (\protect\ref{TopoBerry}) as a function of $%
t_{a}/t_{b}$. It is quantized in the insulators. (b)$\sim $(c) Energy
spectrum as a function of $t_{a}/t_{b}$ for (b) the bulk, (c) a nanoribbon
with width 64, and (d) a triangle with size 20. In (b), the blue curves
represent the bulk energy (\protect\ref{EqK}) at the $K$ and $K^{\prime }$
points, and the cyan lines represent the bulk energy (\protect\ref{EqG}) at
the $\Gamma $ point, which are analytically obtained. In the topological
phase, gapless edge states are absent in nanoribbon geometry as in (c), but
zero-energy corner states emerge in triangle geometry as indicated by the
magenta flat-line segment in (d). The segment slightly deviates from the
region $(-1,1/2)$ due to the finite size effect. }
\label{FigGap}
\end{figure}

\subsection{Clock symmetry}

The Hamiltonian (\ref{H3}) has $\mathbb{Z}_{3}$ clock symmetry\cite{Fend},%
\begin{equation}
\tau H\left( \mathbf{k}\right) \tau ^{\dagger }=\omega H\left( R\mathbf{k}%
\right) ,
\end{equation}%
where $R$ rotates the momentum by 120 degrees as%
\begin{align}
R\left( k_{x},0\right) & =\left( -\frac{k_{x}}{2},\frac{\sqrt{3}k_{y}}{2}%
\right) , \\
R\left( -\frac{k_{x}}{2},\frac{\sqrt{3}k_{y}}{2}\right) & =\left( -\frac{%
k_{x}}{2},-\frac{\sqrt{3}k_{y}}{2}\right) , \\
R\left( -\frac{k_{x}}{2},-\frac{\sqrt{3}k_{y}}{2}\right) & =\left(
k_{x},0\right) ,
\end{align}%
making the energy spectrum have $\mathbb{Z}_{3}$ symmetry in the complex
plane as in Eq.(\ref{En}). In addition, there is an anti-unitary symmetry 
\begin{equation}
KH\left( \mathbf{k}\right) K=H^{\ast }\left( \mathbf{k}\right) ,
\end{equation}
where $K$ implies complex conjugate. It leads to reflection symmetry between 
$E$ and $E^{\ast }$. As a result, the energy spectrum has $C_{3v}$ symmetry
in the complex plane, which consists of the three-fold rotational symmetry
and three reflection symmetries.

\begin{figure}[t]
\centerline{\includegraphics[width=0.48\textwidth]{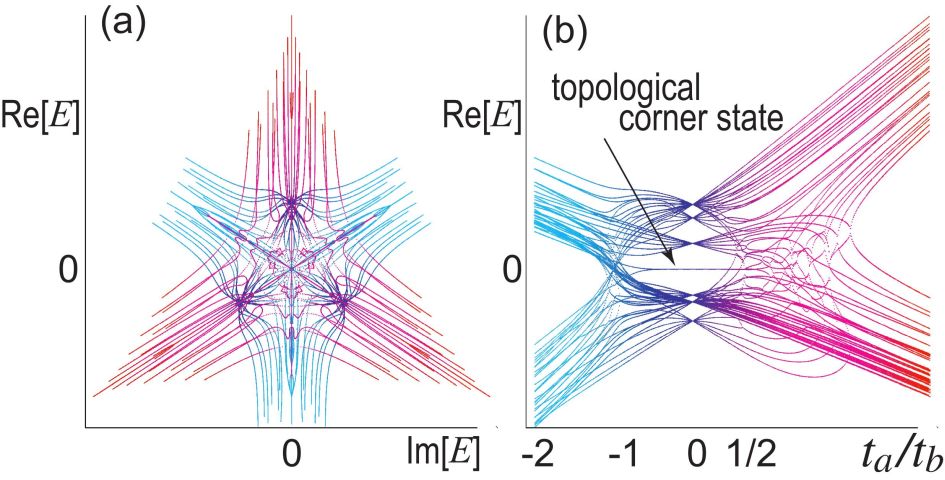}}
\caption{ Energy spectrum of a triangle, where (a) the vertical axis is Re[$E
$] and the horizontal axis is Im[$E$], while (b) the horizontal axis is $%
t_{a}/t_{b}$. Color indicates the value of $t_{a}/t_{b}$, where the color
pallet is the same as in (b). $C_{3v}$ symmetry in the complex energy plane
is manifest in (a). The emergence of zero-energy corner states is clear in
(b). We have used a triangle with size $6$.}
\label{FigColor}
\end{figure}

The system also has a generalized chiral symmetry for a three-band model\cite%
{Ni}, 
\begin{eqnarray}
\sigma H\sigma ^{-1} &=&H_{1},\qquad \sigma H_{1}\sigma ^{-1}=H_{2},  \notag
\\
H+H_{1}+H_{2} &=&0.
\end{eqnarray}
We show the bulk energy spectrum in Fig.\ref{FigColor}(a) and Fig.\ref%
{FigSpat}(a), where $C_{3v}$ symmetry is manifest for all parameters.

\subsection{Phase diagram}

The notion of insulator and metal is generalized to the non-Hermitian
Hamiltonian in two ways. On is a point-gap insulator\cite{Gong,Kawabata},
where $|E|$ has a gap. The other is a line-gap insulator\cite{Gong,Kawabata}%
, where Re$\left[ E\right] $ or Im$\left[ E\right] $ has a gap. In our
model, we adopt the definition of the point-gap insulator due to $\mathbb{Z}%
_{3}$ symmetry. We are able to determine the energy spectrum analytically at
the $\Gamma =(0,0)$ point as%
\begin{equation}
E\left( 0,0\right) =(t_{a}+t_{b}),\quad \omega (t_{a}+t_{b}),\quad \omega
^{2}(t_{a}+t_{b}),  \label{EqG}
\end{equation}%
and at the $K=(4\pi /3,0)$ and $K^{\prime }=(-4\pi /3,0)$ points as%
\begin{equation}
E^{3}(\pm 4\pi /3,0)=\left( t_{a}+t_{b}\right) \left( t_{a}-2t_{b}\right)
\left( 2t_{a}-t_{b}\right) .  \label{EqK}
\end{equation}%
The point gap closes at the $K$ and $K^{\prime }$ points for $%
t_{a}/t_{b}=1/2 $, $t_{a}/t_{b}=2$, and at the $K$, $K^{\prime }$ and $%
\Gamma $ points for $t_{a}/t_{b}=-1$, as in Fig.\ref{FigGap}(b). This is
also confirmed numerically by calculating the band spectrum as in Fig.\ref%
{FigGap}(b).

\begin{figure*}[t]
\centerline{\includegraphics[width=0.98\textwidth]{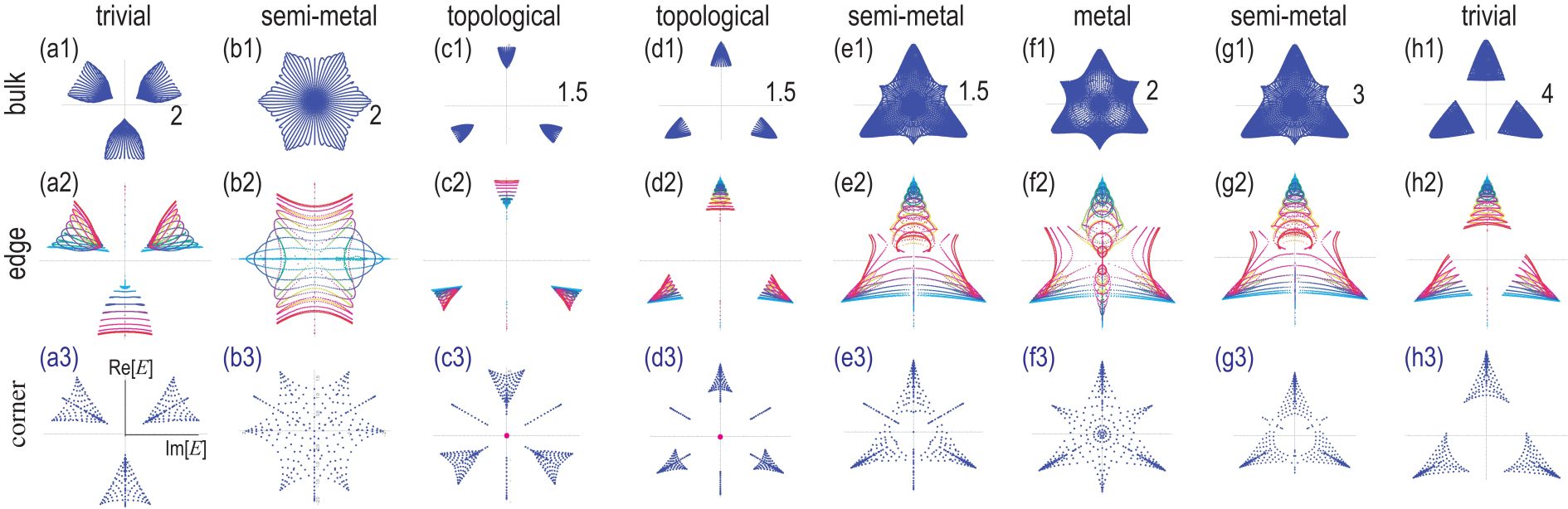}}
\caption{Complex energy spectrum (a1)$\sim$ (h1) for the bulk, (a2)$\sim$
(h2) for a nanoribbon and (a3)$\sim$ (h3) for a triangle, where the hopping
parameters are (a1)$\sim$ (a3) $t_{a}=-1.5t_{b}$, (b1)$\sim$ (b3) $%
t_{a}=-t_{b}$, (c1)$\sim$ (c3) $t_{a}=-0.25t_{b}$, (d1)$\sim$ (d3) $%
t_{a}=0.25t_{b}$, (e1)$\sim$ (e3) $t_{a}=0.5t_{b}$, (f1)$\sim$ (f3) $%
t_{a}=1.25t_{b}$, (g1)$\sim$ (g3) $t_{a}=2t_{b}$ and (h1)$\sim$ (h3) $%
t_{a}=2.5t_{b}$. The horizontal axis is Im[$E$] and the vertical axis is Re[$%
E $]. The numerical value on the horizontal axis in (a1)$\sim$ (h1) is the
energy in unit of $t_b$. Color indicates the momentum of along the
nanoribbon direction, where red color indicates $k=\protect\pi $ and blue
color indicates $k=0$ in a nanoribbon. We have used a nanoribbon with width $%
128$ for (a2)$\sim$ (h2), and a triangle with size $16$ for (a3)$\sim$ (h3).
The magenta dots in (c3) and (d3) represent the topological corner states.}
\label{FigSpat}
\end{figure*}

\begin{figure*}[t]
\centerline{\includegraphics[width=0.98\textwidth]{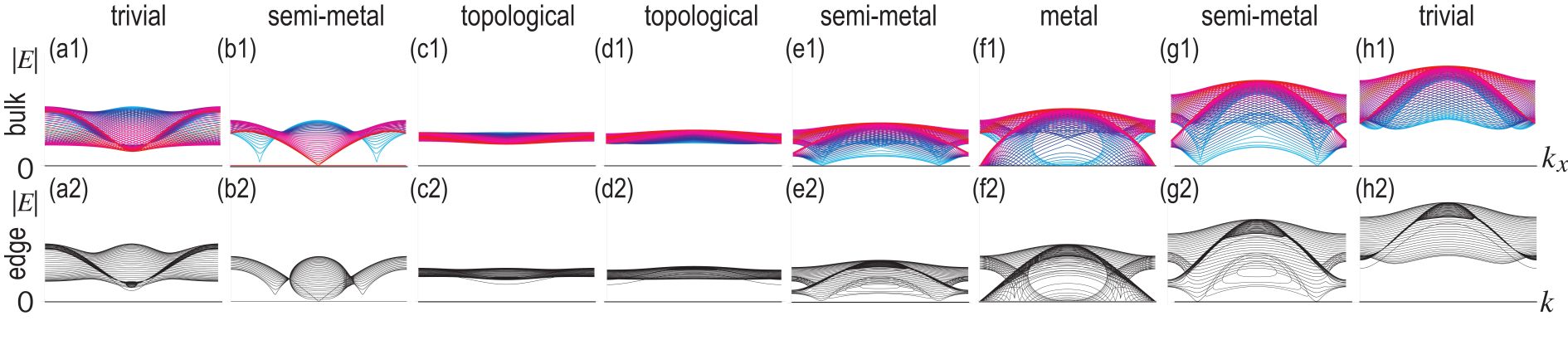}}
\caption{(a1)$\sim$(h1) Bulk band structure, where the horizontal axis is $%
k_{x}$. Color in the spectrum indicates $k_{y}$, where red color indicates $%
k_y=0$ while cyan color indicates $k_y=2\protect\pi /\protect\sqrt{3}$. (a2)$%
\sim$(h2) Band structure of a nanoribbon, where the horizontal axis is $k$.
The vertical axis is $|E|$. The values of $t_{a}$ and $t_{b}$ are the same
as in Fig.\protect\ref{FigSpat}. It is notable that there are no gapless
edge states in the topological phase.}
\label{FigRibbon}
\end{figure*}

\subsection{Topological number}

The topological number is given by the Berry phase defined by%
\begin{equation}
Q\equiv \frac{1}{2\pi i}\int_{0}^{2\pi }\left\langle \psi _{0}\left(
k_{x},0\right) \right\vert \partial _{k_{x}}\left\vert \psi _{0}\left(
k_{x},0\right) \right\rangle dk_{x},  \label{TopoBerry}
\end{equation}%
where $\psi _{0}$ is the eigen function of the Hamiltonian (\ref{H3}) for
the bulk, whose eigen energy is real along the $k_{x}$ axis (i.e., $k_{y}=0$%
). We calculate it numerically, whose results are shown in Fig.\ref{FigGap}%
(a). We find $Q=1$ for $-1<t_{a}/t_{b}<1/2$, and $Q=0$ for $t_{a}/t_{b}<-1$
and $t_{a}/t_{b}>2$, while it continuously changes from $1$ to $0$ for $%
1/2<t_{a}/t_{b}<2$. In fact, $Q$ is quantized in the insulator phases.

\subsection{Edge states}

We calculate the energy spectrum in a nanoribbon numerically. Edge states
are observed in Fig.\ref{FigSpat}(a2)$\sim $(h2), where the complex energy
spectrum is shown for various momentum $k$ specified by color. Three-fold
symmetry is slightly broken in nanoribbon geometry. It is due to the finite
size effect of a nanoribbon.

We present the band structure of a nanoribbon in Fig.\ref{FigRibbon}(a2)$%
\sim $(h2), where we observe the absence of gapless edge states in the
topological phase.

\subsection{Corner states}

We calculate the energy spectrum in triangle geometry numerically. $C_{3v}$
symmetry is manifest as shown in Fig.\ref{FigColor}(a) and in Fig.\ref%
{FigSpat}(a3)$\sim $(h3). It is because the triangle respects $\mathbb{Z}%
_{3} $ clock symmetry. We find zero-energy states in the region $%
-1<t_{a}/t_{b}<1/2$, as indicated by a magenta line in Fig.\ref{FigGap}(d).
We also show the energy spectrum as a function of $t_{a}/t_{b}$ in Fig.\ref%
{FigColor}(b), where the emergence of the zero-energy states is manifest in
the topological phase.

Consequently, the present model is a second-order topological insulator in
the region $-1<t_{a}/t_{b}<1/2$, being characterized by the emergence of
topological corner states.

It is possible to obtain explicitly the wave functions of the three corner
states by numerical calculation. We have numerically confirmed that they
satisfy the relations (\ref{TauP}) and (\ref{SigmaP}). Therefore, they are $%
\mathbb{Z}_{3}$ parafermions.

\subsection{Electric-circuit implementation}

Electric circuits are governed by the Kirchhoff current law. By making the
Fourier transformation with respect to time, the Kirchhoff current law is
expressed as 
\begin{equation}
I_{a}\left( \omega \right) =\sum_{b}J_{ab}\left( \omega \right) V_{b}\left(
\omega \right) ,
\end{equation}
where $I_{a}$ is the current between node $a$ and the ground, while $V_{b}$
is the voltage at node $b$. The matrix $J_{ab}\left( \omega \right) $ is
called the circuit Laplacian. Once the circuit Laplacian is given, we can
uniquely setup the corresponding electric circuit. By equating it with the
Hamiltonian $H$ as\cite{TECNature,ComPhys} 
\begin{equation}
J_{ab}\left( \omega \right) =i\omega H_{ab}\left( \omega \right) ,
\label{CircuitLap}
\end{equation}%
it is possible to simulate various topological phases of the Hamiltonian by
electric circuits\cite%
{TECNature,ComPhys,Hel,Lu,YLi,EzawaTEC,EzawaLCR,EzawaSkin,Garcia,Hofmann,EzawaMajo,Tjunc}%
. The relations between the parameters in the Hamiltonian and in the
electric circuit are determined by this formula.

The circuit Laplacian is constructed as follows. To simulate the positive
and negative hoppings in the Hamiltonian, we replace them with the
capacitance $i\omega C$ and the inductance $1/i\omega L$, respectively. We
note that $\sin k=(e^{ik}-e^{-ik})/2i$ represents an imaginary hopping in
the tight-bind model. The imaginary hopping is realized by an operational
amplifier\cite{Hofmann}.

We thus make the following replacements with respect to hoppings in the
Hamiltonian to derive the circuit Laplacian: (i) $+X\rightarrow i\omega
C_{X} $ for $X=t_{a}$ and $t_{b}$, where $C_{X}$ represents the capacitance
whose value is $X$ [pF]. (ii) $-X\rightarrow 1/i\omega L_{X}$ for $X=t_{a}/2$
and $t_{b}/2$, where $L_{X}$ represents the inductance whose value is $X$ [$%
\mu $H].

\begin{figure}[t]
\centerline{\includegraphics[width=0.49\textwidth]{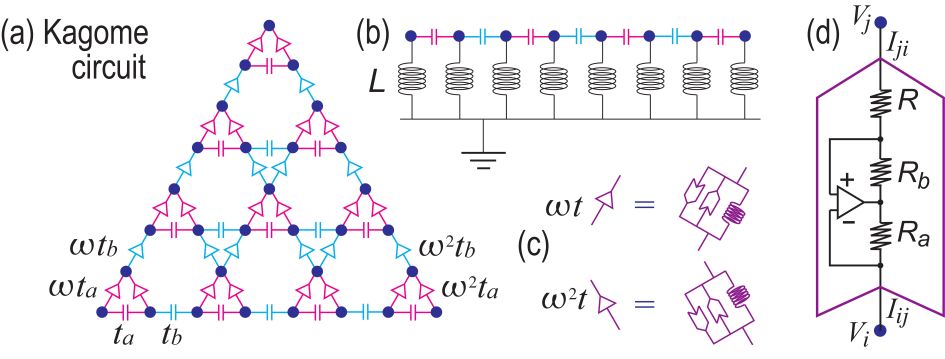}}
\caption{(a) Illustration of the breathing Kagome circuit corresponding to
Fig.\protect\ref{FigLattice}(a). (b) Each node is grounded by an inductor.
(c) Complex hoppings $\propto e^{\pm 2\protect\pi i/3}$ are realized by a
parallel connection of an inductor and an operational amplifier. We show a
triangle with size $4$.}
\label{FigKagomeCircuit}
\end{figure}

We explicitly study the breathing Kagome lattice described by (\ref{H3}),
where the electric circuit is given by Fig.\ref{FigKagomeCircuit}(a). The
Hamiltonian (\ref{H3}) is decomposed into%
\begin{equation}
H=H_{1}+H_{2},
\end{equation}%
with 
\begin{equation}
H_{1}=\left( 
\begin{array}{ccc}
0 & h_{12} & -\frac{1}{2}h_{13} \\ 
h_{12}^{\ast } & 0 & \omega ^{2}h_{23} \\ 
-\frac{1}{2}h_{13}^{\ast } & \omega ^{2}h_{23}^{\ast } & 0%
\end{array}%
\right)
\end{equation}%
and%
\begin{equation}
H_{2}=\frac{\sqrt{3}}{2}\left( 
\begin{array}{ccc}
0 & 0 & ih_{13} \\ 
0 & 0 & -ih_{23} \\ 
ih_{13}^{\ast } & -ih_{23}^{\ast } & 0%
\end{array}%
\right) ,
\end{equation}%
where $H_{1}$ is Hermitian ($H_{1}^{\dagger }=H_{1}$),\ and $H_{2}$ is
anti-Hermitian ($H_{2}^{\dagger }=-H_{2}$). It is necessary to construct
imaginary hopping Hamiltonians%
\begin{equation}
\frac{\sqrt{3}}{2}t_{a}\left( 
\begin{array}{ccc}
0 & 0 & i \\ 
0 & 0 & 0 \\ 
i & 0 & 0%
\end{array}%
\right) ,  \label{EqA}
\end{equation}%
and%
\begin{equation}
\frac{\sqrt{3}}{2}t_{a}\left( 
\begin{array}{ccc}
0 & 0 & 0 \\ 
0 & 0 & -i \\ 
0 & -i & 0%
\end{array}%
\right)  \label{EqB}
\end{equation}%
for the magenta lines in Fig.\ref{FigKagomeCircuit}(a). They are constructed
by using operational amplifiers and resistors.

\begin{figure}[t]
\centerline{\includegraphics[width=0.49\textwidth]{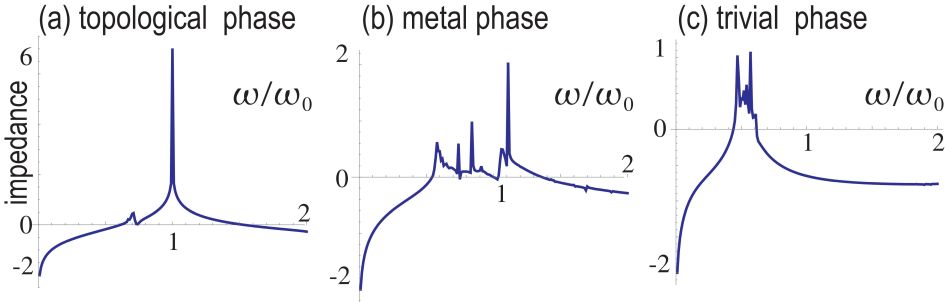}}
\caption{Impedance as a function of $\protect\omega $. The vertical axis is $%
\log _{10}Z$ and the horizontal axis is $\protect\omega /{\protect\omega _{0}%
}$. (a) topological phase with $t_{a}/t_{b}=0.25$, (b) metallic phase with $%
t_{a}/t_{b}=1$ and (c) trivial phase with $t_{a}/t_{b}=2.5$. A prominent
peak is found in the topological phase. }
\label{FigImpe}
\end{figure}

We review a negative impedance converter with current inversion based on an
operational amplifier with resistors\cite{Hofmann}. The voltage-current
relation for the operational amplifier circuit is given by\cite{Hofmann} 
\begin{equation}
\left( 
\begin{array}{c}
I_{1} \\ 
I_{2}%
\end{array}%
\right) =\frac{1}{R}\left( 
\begin{array}{cc}
-\nu & \nu \\ 
-1 & 1%
\end{array}%
\right) \left( 
\begin{array}{c}
V_{1} \\ 
V_{2}%
\end{array}%
\right) ,
\end{equation}%
with $\nu =R_{b}/R_{a}$, where $R$, $R_{a}$ and $R_{b}$ are the resistances
in an operational amplifier: See Fig.\ref{FigKagomeCircuit}(d). We note that
the resistors in the operational amplifier circuit are tuned to be $\nu =1$
in the literature\cite{Hofmann} so that the system becomes Hermitian, where
the corresponding Hamiltonian represents a spin-orbit interaction.

In this paper, we use two negative impedance converters parallelly connected
with the opposite direction as in Fig.\ref{FigKagomeCircuit}(c). The circuit
Laplacian due to these two converters is given by%
\begin{equation}
\frac{1}{R}\left[ \left( 
\begin{array}{cc}
-\nu & \nu \\ 
-1 & 1%
\end{array}%
\right) +\left( 
\begin{array}{cc}
1 & -1 \\ 
\nu & -\nu%
\end{array}%
\right) \right] =\frac{1}{R}\left( 
\begin{array}{cc}
1-\nu & \nu -1 \\ 
\nu -1 & 1-\nu%
\end{array}%
\right) .
\end{equation}%
It corresponds to the Hamiltonian%
\begin{equation}
H=\frac{1}{i\omega R}\left( 
\begin{array}{cc}
1-\nu & \nu -1 \\ 
\nu -1 & 1-\nu%
\end{array}%
\right) .
\end{equation}%
It is embedded in the 3$\times $3 matrix as%
\begin{equation}
H=\frac{1}{i\omega R}\left( 
\begin{array}{ccc}
1-\nu & 0 & \nu -1 \\ 
0 & 0 & 0 \\ 
\nu -1 & 0 & 1-\nu%
\end{array}%
\right) ,  \label{EqAA}
\end{equation}%
where we have set%
\begin{equation}
\frac{\sqrt{3}}{2}t_{a}=\frac{1-\nu }{\omega R}
\end{equation}%
with $\nu <1$, and%
\begin{equation}
H=\frac{1}{i\omega R}\left( 
\begin{array}{ccc}
0 & 0 & 0 \\ 
0 & 1-\nu & \nu -1 \\ 
0 & \nu -1 & 1-\nu%
\end{array}%
\right) ,  \label{EqBB}
\end{equation}%
where we have set%
\begin{equation}
\frac{\sqrt{3}}{2}t_{a}=\frac{\nu -1}{\omega R}
\end{equation}%
with $\nu >1$. These matrices are different from Eqs.(\ref{EqA}) and (\ref%
{EqB}) by the diagonal terms. They are cancelled by adding a resistor (for $%
\nu >1$)\ or an operational amplifier (for $\nu <1$)\ with the amount of%
\begin{equation}
\frac{1-\nu }{i\omega R}
\end{equation}%
between a lattice site and the ground.

\subsection{Impedance resonance}

The zero-energy parafermion corner states are well observed by impedance
resonance, which is defined\cite{Hel} by 
\begin{equation}
Z_{ab}=V_{a}/I_{b}=G_{ab}
\end{equation}
where $G=J^{-1}$ is the Green function. It diverges at the frequency where
the admittance is zero ($J=0$). Taking the nodes $a$ and $b$ at two corners,
we show the impedance in topological, metallic and trivial phases in Figs.%
\ref{FigImpe}(a)$\sim $(c), respectively. A strong impedance peak is
observed at the critical frequency $\omega _{0}\equiv 1/\sqrt{LC}$ only in
the topological phase. It signals the emergence of zero-energy parafermion
corner states.

\begin{figure}[t]
\centerline{\includegraphics[width=0.48\textwidth]{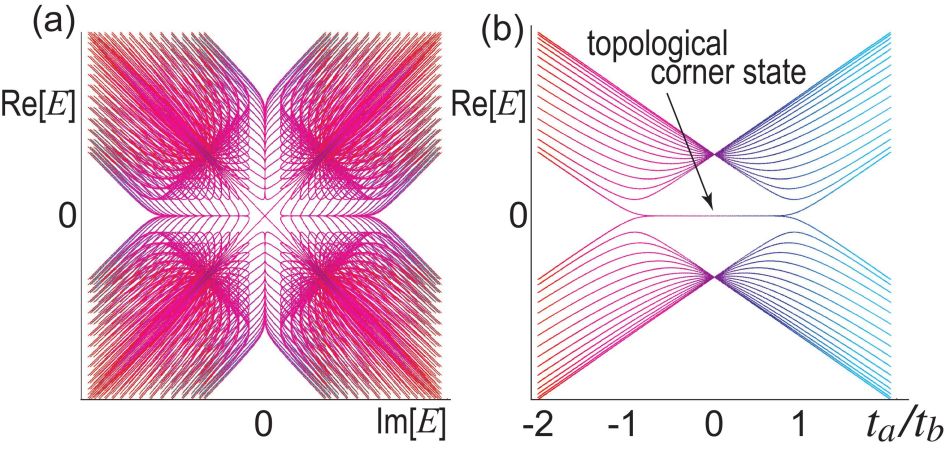}}
\caption{Energy spectrum of a square, where (a) the vertical axis is Re[$E$]
and the horizontal axis is Im[$E$], while (b) the horizontal axis is $%
t_{a}/t_{b}$. Color indicates the value of $t_{a}/t_{b}$, where the color
pallet is the same as in (b). $C_{4v}$ symmetry in the complex energy plane
is manifest in (a). The emergence of zero-energy corner states is clear in
(b). }
\label{FigColorQ}
\end{figure}

\section{$\mathbb{Z}_{4}$ Parafermion}

We proceed to a model with $d=4$. $\mathbb{Z}_{4}$ parafermions are
represented by the shift operator\cite{Zohar,Fend,AliceaPara}%
\begin{equation}
\gamma _{1}\equiv \tau =\left( 
\begin{array}{cccc}
0 & 0 & 0 & 1 \\ 
1 & 0 & 0 & 0 \\ 
0 & 1 & 0 & 0 \\ 
0 & 0 & 1 & 0%
\end{array}%
\right) ,
\end{equation}%
and the clock operator\cite{Zohar,Fend,AliceaPara}%
\begin{equation}
\gamma _{2}\equiv \sigma =\text{diag.}\left( 1,i,-1,-i\right) .
\end{equation}%
Here, $\tau $ and $\sigma $ satisfy the $\mathbb{Z}_{4}$ parafermion
relations,%
\begin{equation}
\tau ^{4}=\sigma ^{4}=1,\qquad \tau \sigma =\omega \sigma \tau .
\end{equation}%
In the $\mathbb{Z}_{4}$ clock-symmetric model, the energy spectrum is
composed of quartets $E_{n}^{\left( 0,1,2,3\right) }$, $n=0,1,2,\cdots $, 
\begin{equation}
E_{n}^{\left( 0,1,2,3\right) }=\varepsilon _{n},\quad i\varepsilon
_{n},\quad -\varepsilon _{n},\quad -i\varepsilon _{n}.  \label{Enn}
\end{equation}%
The system is necessarily non-Hermitian because the eigen energies are
complex except for zero-energy states.

\subsection{Zero-energy parafermion states}

It follows from Eq.(\ref{Enn}) that only the zero-energy states form a set
of degenerate states respecting $\mathbb{Z}_{4}$ clock symmetry. They are $%
\mathbb{Z}_{4}$ parafermion states. We denote them as $\left\vert \psi
_{0}\right\rangle $, $\left\vert \psi _{1}\right\rangle $, $\left\vert \psi
_{2}\right\rangle $ and $\left\vert \psi _{3}\right\rangle $. They are
characterized by the properties 
\begin{align}
\tau \left\vert \psi _{0}\right\rangle & =\left\vert \psi _{1}\right\rangle
,\quad \tau \left\vert \psi _{1}\right\rangle =\left\vert \psi
_{2}\right\rangle , \\
\tau \left\vert \psi _{2}\right\rangle & =\left\vert \psi _{3}\right\rangle
,\quad \tau \left\vert \psi _{3}\right\rangle =\left\vert \psi
_{0}\right\rangle , \\
\sigma \left\vert \psi _{0}\right\rangle & =\left\vert \psi
_{0}\right\rangle ,\quad \sigma \left\vert \psi _{1}\right\rangle
=i\left\vert \psi _{1}\right\rangle , \\
\sigma \left\vert \psi _{2}\right\rangle & =-\left\vert \psi
_{2}\right\rangle ,\quad \sigma \left\vert \psi _{3}\right\rangle
=-i\left\vert \psi _{3}\right\rangle ,
\end{align}%
from which the matrix representations (\ref{Tau}) and (\ref{DigW}) follow.
Then, the $\mathbb{Z}_{4}$ parafermion relations (\ref{EqPara}) are
verified. Namely, it is necessary and sufficient to examine Eqs.(\ref{TauP})
and (\ref{SigmaP}) for a triplet set of zero-energy states in order to show
that they are $\mathbb{Z}_{4}$ parafermions.

\subsection{Breathing square lattice}

The quadrupole insulator has been proposed on the breathing square lattice%
\cite{Science}. We propose a model possessing $\mathbb{Z}_{4}$ parafermions
on the breathing square lattice. The bulk Hamiltonian is given by%
\begin{equation}
H=\left( 
\begin{array}{cccc}
0 & f_{x}^{\ast } & 0 & if_{y}^{\ast } \\ 
-f_{x} & 0 & if_{y}^{\ast } & 0 \\ 
0 & -if_{y} & 0 & -f_{x} \\ 
-if_{y} & 0 & f_{x}^{\ast } & 0%
\end{array}%
\right) ,  \label{H4}
\end{equation}%
with 
\begin{equation}
f_{x}=t_{a}+t_{b}e^{ik_{x}},\qquad f_{y}=t_{a}+t_{b}e^{ik_{y}},
\end{equation}
where we have introduced two hopping parameters $t_{a}$ and $t_{b}$, which
are shown in Fig.\ref{FigLattice}(b4).

The Hamiltonian (\ref{H4}) has $\mathbb{Z}_{4}$ clock symmetry\cite{Fend},%
\begin{equation}
\tau H\left( \mathbf{k}\right) \tau ^{\dagger }=-iH\left( R\mathbf{k}\right)
,
\end{equation}%
where $R$ rotates the momentum by $90$ degrees as%
\begin{align}
R\left( k_{x},0\right) & =\left( 0,-k_{y}\right) , \\
R\left( 0,k_{y}\right) & =\left( k_{x},0\right) ,
\end{align}%
making the energy spectrum have $\mathbb{Z}_{4}$ symmetry in the complex
plane as in Eq.(\ref{Enn}).

\subsection{Edge and corner states}

The topological number is defined by (\ref{TopoBerry}), where $\psi _{0}$\
is the eigen function of the Hamiltonian (\ref{H4}) for the bulk. We find
the topological insulator phase for $\left\vert t_{a}/t_{b}\right\vert <1$,
where $Q=1$\ and the trivial insulator phase for $\left\vert
t_{a}/t_{b}\right\vert >1$, where $Q=0$.

We calculate the energy spectrum in square geometry numerically. $C_{4v}$\
symmetry is manifest as shown in Fig.\ref{FigColorQ}(a). It is because the
square lattice respects $\mathbb{Z}_{4}$ clock symmetry. We also show the
energy spectrum as a function of $t_{a}/t_{b}$\ in Fig.\ref{FigColorQ}(b),
where the emergence of the zero-energy states is manifest in the topological
phase.

Consequently, the present model is a second-order topological insulator in
the region $|t_{a}/t_{b}|<1$, being characterized by the emergence of
topological corner states.

It is possible to obtain numerically the wave functions of the four corner
states. We have confirmed that they satisfy the relations (\ref{TauP}) and (%
\ref{SigmaP}). Therefore, they are $\mathbb{Z}_{4}$\ parafermions.

\begin{figure}[t]
\centerline{\includegraphics[width=0.49\textwidth]{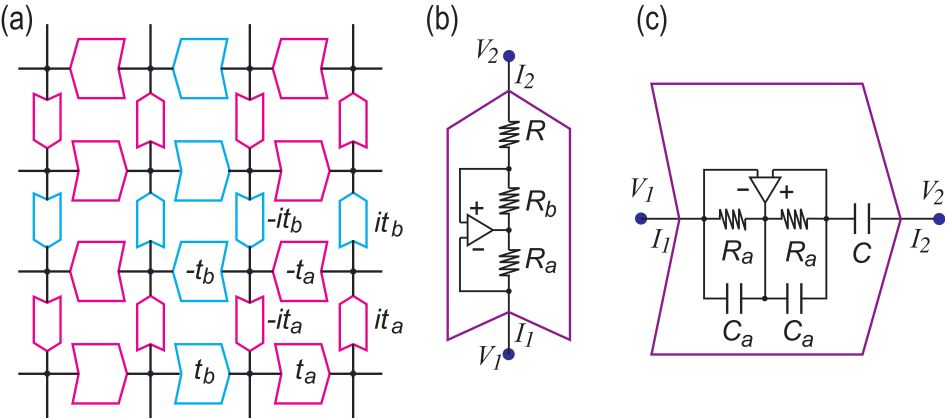}}
\caption{(a) Electric circuit implementation of the breathing square model
shown in Fig.\protect\ref{FigLattice}(b), where the hopping parameters
correspond to those in Fig.\protect\ref{FigLattice}(b4). (b) Configuration
of negative impedance converter with current inversion realizing imaginary
hopping\protect\cite{Hofmann}, and (c) the one realizing nonreciprocal
hopping\protect\cite{NonR}, corresponding to those in (a). }
\label{FigOpeamp}
\end{figure}

\subsection{Electric-circuit implementation}

In the Hamiltonian (\ref{H4}), there are nonreciprocal hopping terms along
the $x$ axis. Nonreciprocal hopping is constructed by a combination of
operational amplifier and capacitors\cite{NonR},%
\begin{equation}
\left( 
\begin{array}{c}
I_{ij} \\ 
I_{ji}%
\end{array}%
\right) =i\omega C\left( 
\begin{array}{cc}
-1 & 1 \\ 
-1 & 1%
\end{array}%
\right) \left( 
\begin{array}{c}
V_{i} \\ 
V_{j}%
\end{array}%
\right) ,
\end{equation}%
It corresponds to the Hamiltonian%
\begin{equation}
H=C\left( 
\begin{array}{cc}
-1 & 1 \\ 
-1 & 1%
\end{array}%
\right) .
\end{equation}%
We add an inductor or a capacitor in order to cancel the diagonal term.

We construct an electric circuit as shown in Fig.\ref{FigOpeamp}(a), where
the $x$ axis is constructed by Fig.\ref{FigOpeamp}(c) and the $y$ axis is
constructed by Fig.\ref{FigOpeamp}(b).

\section{$\mathbb{Z}_{6}$ Parafermion}

Finally, we construct a $\mathbb{Z}_{6}$ parafermion model. The parafermion
operators are represented by the shift operator\cite{Zohar,Fend,AliceaPara}%
\begin{equation}
\gamma _{1}\equiv \tau =\left( 
\begin{array}{cccccc}
0 & 0 & 0 & 0 & 0 & 1 \\ 
1 & 0 & 0 & 0 & 0 & 0 \\ 
0 & 1 & 0 & 0 & 0 & 0 \\ 
0 & 0 & 1 & 0 & 0 & 0 \\ 
0 & 0 & 0 & 1 & 0 & 0 \\ 
0 & 0 & 0 & 0 & 1 & 0%
\end{array}%
\right) ,
\end{equation}%
and the clock operator\cite{Zohar,Fend,AliceaPara}%
\begin{equation}
\gamma _{2}\equiv \sigma =\text{diag.}\left( 1,\omega ,\omega ^{2},\omega
^{3},\omega ^{4},\omega ^{5}\right) .
\end{equation}%
Here, $\tau $ and $\sigma $ satisfy the $\mathbb{Z}_{6}$ parafermion
relations,%
\begin{equation}
\tau ^{6}=\sigma ^{6}=1,\qquad \tau \sigma =\omega \sigma \tau .
\end{equation}%
In the $\mathbb{Z}_{6}$ clock-symmetric model, the energy spectrum is
composed of sextets $E_{n}^{\left( 0,1,2,3,4,5\right) }$, $n=0,1,2,\cdots $,

\begin{equation}
E_{n}^{\left( 0,1,2,3,4,5\right) }=\varepsilon _{n},\quad \omega \varepsilon
_{n},\quad \omega ^{2}\varepsilon _{n},\quad \omega ^{3}\varepsilon
_{n},\quad \omega ^{4}\varepsilon _{n},\quad \omega ^{5}\varepsilon _{n}.
\end{equation}%
The system is necessarily non-Hermitian because the eigen energies are
complex except for zero-energy states.

\begin{figure}[t]
\centerline{\includegraphics[width=0.48\textwidth]{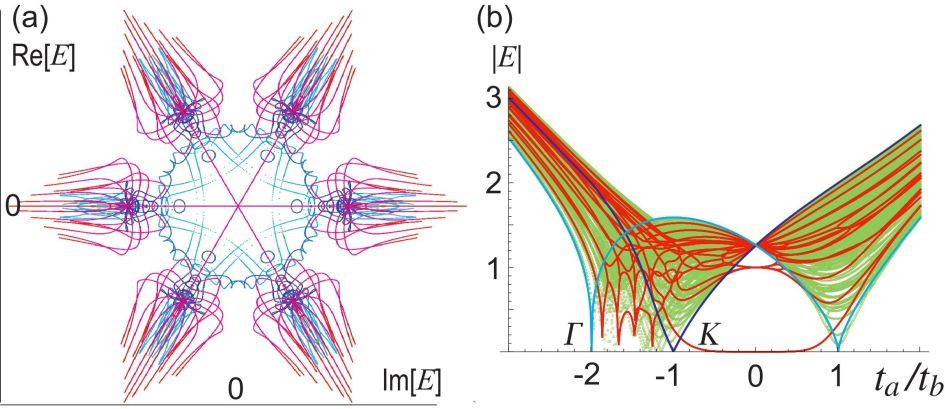}}
\caption{Energy spectrum of a hexagon shown in Fig.\protect\ref{FigLattice}%
(c2). (a) the vertical axis is Re[$E$] and the horizontal axis is Im[$E$],
while (b) the vertical axis is $|E|$ and the horizontal axis is $t_{a}/t_{b}$%
. Color indicates the value of $t_{a}/t_{b}$ in (a). $C_{6v}$ symmetry in
the complex energy plane is manifest in (a). The emergence of zero-energy
corner states is clear in (b). The green curves are calculated from the bulk
Hamiltonian, while red curves are calculated based on the hexagon. The cyan
curve is the energy at the $\Gamma $ point, while the blue curve is the
energy at the $K$ point calculated based on the bulk Hamiltonian.}
\label{FigSpatH}
\end{figure}

A Hermitian second-order topological insulator has been proposed on the
breathing honeycomb lattice\cite{Mizoguchi}. We generalize it to a
non-Hermitian model with $\mathbb{Z}_{6}$ clock symmetry by introducing
complex hoppings. The bulk Hamiltonian is defined on the breathing honeycomb
lattice and given by 
\begin{widetext}
\begin{equation}
H=\left( 
\begin{array}{cccccc}
0 & \omega ^{2}t_{b} & 0 & t_{a}e^{ik_{y}} & 0 & \omega ^{4}t_{b} \\ 
\omega ^{5}t_{b} & 0 & \omega ^{3}t_{b} & 0 & \omega t_{a}e^{i\frac{-\sqrt{3}%
k_{x}+k_{y}}{2}} & 0 \\ 
0 & t_{b} & 0 & \omega ^{4}t_{b} & 0 & \omega ^{2}t_{a}e^{-i\frac{\sqrt{3}%
k_{x}+k_{y}}{2}} \\ 
\omega ^{3}t_{a}e^{-ik_{y}} & 0 & \omega t_{b} & 0 & \omega ^{5}t_{b} & 0 \\ 
0 & \omega ^{4}t_{a}e^{i\frac{\sqrt{3}k_{x}-k_{y}}{2}} & 0 & \omega ^{2}t_{b}
& 0 & t_{b} \\ 
\omega t_{b} & 0 & \omega ^{5}t_{a}e^{i\frac{\sqrt{3}k_{x}+k_{y}}{2}} & 0 & 
\omega ^{3}t_{b} & 0%
\end{array}%
\right) .
\end{equation}%
\end{widetext}

We diagonalize this Hamiltonian for a hexagon shown in Fig.\ref{FigLattice}%
(c2). $C_{6v}$\ symmetry is manifest in the complex energy plane as in Fig.%
\ref{FigSpatH}(a). Furthermore, we find six zero-energy topological corner
states for $\left\vert t_{a}/t_{b}\right\vert <1$, as shown in Fig.\ref%
{FigSpatH}(b). We also find the trivial insulator phase for $t_{a}/t_{b}<-2$
and $t_{a}/t_{b}>1$. Additionally, there is metallic phase for $%
-2<t_{a}/t_{b}<-1$.

\section{Conclusion}

We have constructed a $\mathbb{Z}_{3}$ parafermion model on the breathing
Kagome lattice, a $\mathbb{Z}_{4}$ parafermion model on the breathing square
lattice and a $\mathbb{Z}_{6}$ parafermion model on the breathing honeycomb
lattice. These model exhaust all the possible realization of $\mathbb{Z}_{d}$
parafermions since there are only three-fold, four-fold and six-fold
rotational symmetries that are compatible with the periodic lattices. We
note that two-fold symmetry corresponds to Majorana fermions.

The author is very much grateful to Y. Tanaka and N. Nagaosa for helpful
discussions on the subject. This work is supported by the Grants-in-Aid for
Scientific Research from MEXT KAKENHI (Grants No. JP17K05490 and No.
JP18H03676). This work is also supported by CREST, JST (JPMJCR16F1 and
JPMJCR20T2).

\end{document}